\pdfoutput=1

\documentclass[11pt,twoside,a4paper,cmspaper,final,collab]{cms-tdr}
\def\svnVersion{458639P}\def\cmsCernNoTag{CERN-EP-2017-271}\def\cmsCernDate{\today}\def\cmsMessage{Published in Physics Letters B as \href{http://dx.doi.org/10.1016/j.physletb.2018.04.007}{\doi{10.1016/j.physletb.2018.04.007.}}}

\begin{document}\cmsNoteHeader{EXO-17-002}

\hyphenation{had-ron-i-za-tion}
\hyphenation{cal-or-i-me-ter}
\hyphenation{de-vices}
\RCS$Revision: 458633 $
\RCS$HeadURL: svn+ssh://svn.cern.ch/reps/tdr2/papers/EXO-17-002/trunk/EXO-17-002.tex $
\RCS$Id: EXO-17-002.tex 458633 2018-05-05 18:17:38Z rgarg $
\newcommand{\jet}{\ensuremath{\text{jet}}\xspace}
\newcommand{\jets}{\ensuremath{\text{jets}}\xspace}
\newcommand{\cPqst}{{\ensuremath{\PQq^\star}\xspace}}
\newcommand{\cPbst}{{\ensuremath{\PQb^\star}\xspace}}
\newcommand{\GAMbJET}{\ensuremath{\gamma\mathrm{+}\text{\PQb jet}}\xspace}

\newlength\cmsFigWidth
\ifthenelse{\boolean{cms@external}}{\setlength\cmsFigWidth{0.98\columnwidth}}{\setlength\cmsFigWidth{0.4\textwidth}}
\ifthenelse{\boolean{cms@external}}{\providecommand{\cmsLeft}{upper\xspace}}{\providecommand{\cmsLeft}{left\xspace}}
\ifthenelse{\boolean{cms@external}}{\providecommand{\cmsRight}{lower\xspace}}{\providecommand{\cmsRight}{right\xspace}}
\cmsNoteHeader{EXO-17-002}
\title{Search for excited quarks of light and heavy flavor in $\gamma+\jet$ final states in proton-proton collisions at $\sqrt{s}=13\TeV$}

\date{\today}

\abstract{
A search is presented for excited quarks of light and heavy flavor that decay to
$\gamma+\text{jet}$ final states. The analysis is based on data corresponding
to an integrated luminosity of 35.9\fbinv collected by the CMS
experiment in proton-proton collisions at $\sqrt{s}=13\TeV$ at the LHC.
A signal would appear as a resonant contribution to the invariant mass spectrum
of the $\gamma+\text{jet}$ system, above the background expected from standard model processes.
No resonant excess is found, and
upper limits are set on the product of the excited quark cross section and its
branching fraction as a function of its mass. These are the most stringent limits to date in the $\gamma\mathrm{+jet}$ final state, and
exclude excited light quarks with masses below 5.5\TeV and excited b quarks
with masses below 1.8\TeV, assuming standard model like coupling strengths.
}

\hypersetup{
pdfauthor={CMS Collaboration},
pdftitle={Search for excited states of light and heavy flavor quarks in the gamma+jet final state in proton-proton collisions at sqrt(s) = 13 TeV},
pdfsubject={CMS},
pdfkeywords={CMS, physics, excited quarks}
}

\maketitle

\section{Introduction}
High energy proton-proton collisions resulting in a photon and a jet with large transverse momenta
(\pt) provide a powerful means of searching for new physics. For example, models involving
compositeness~\cite{compositeness_theories2, Baur:1987ga, Baur1989kv} predict excited states of quarks that can be
identified by searching for events that contain a photon and a jet from their decays. We present a search for excited states of light (u,d) and
heavy (b) quarks using this decay signature.

We assume that the coupling between the excited quark ($\cPqst$),
the ordinary quarks, and gauge bosons proceeds through a gauge-invariant magnetic-moment operator, described by the
effective Lagrangian~\cite{Bhattacharya:2009xg}:
\ifthenelse{\boolean{cms@external}}{
\begin{multline}
\mathcal{L}_\text{int} = \frac{1}{2 \, \Lambda}\overline{\cPqst}_{\text{R}} \, \sigma^{\mu\nu}
\Big[g_{s}f_{s}\frac{\lambda_{a}}{2}G^{a}_{\mu\nu}\\
+gf\frac{\tau}{2} \cdot W_{\mu\nu}\;+\;g'f'\frac{Y}{2}B_{\mu\nu} \Big] {\cPq}_{\text{L}} + \text{h.c.},
\label{eq:Lagrangian}
\end{multline}
}{
\begin{equation}
\mathcal{L}_\text{int} = \frac{1}{2 \, \Lambda}\overline{\cPqst}_{\text{R}}\, \sigma^{\mu\nu}
\left[g_{s}f_{s}\frac{\lambda_{a}}{2}G^{a}_{\mu\nu}\;+\;gf\frac{\tau}{2} \cdot W_{\mu\nu}\;+\;g'f'\frac{Y}{2}B_{\mu\nu} \right] {\cPq}_{\text{L}} + \text{h.c.},
\label{eq:Lagrangian}
\end{equation}
}

where $\cPqst_{\text{R}}$ is the right-handed excited quark field; $\sigma_{\mu\nu}$ the Pauli spin matrix; $\cPq_{\text{L}}$ the left-handed quark field;
$G^{a}_{\mu\nu}$, $W_{\mu\nu}$, and $B_{\mu\nu}$ are the field tensors of the
SU(3), SU(2), and U(1) gauge fields respectively;
$\lambda_{a}$, $\tau$, and $Y$ are the corresponding gauge structure constants, and
$g_{s}$, $g$, and $g'$ are the gauge couplings.
The compositeness scale $\Lambda$ is the energy scale typical for these interactions. The quantities $f_{s}$, $f$, and $f'$ are unknown dimensionless
constants that represent the strengths of the excited quark couplings to the standard model (SM) partners. Their values are determined by the compositeness
dynamics, and are usually assumed to be of order unity.

In pp collisions, excited quarks are expected to be produced predominantly through quark-gluon fusion ($\cPq\cPg$), and then decay into
a quark and a gauge boson ($\cPg, \PW, \PZ, \cPgg$). Searches have been performed in different channels
~\cite{Abe:1993sv, Abe:1997hm, Aaltonen:2008dn, Khachatryan:2014aka, Khachatryan:2015sja, Khachatryan:2015dcf, Aad:2015ywd, Aaboud:2017nak},
but no evidence for the existence of
excited quarks has yet been found. This analysis looks for evidence of $\cPq\cPg \to \cPqst \to \cPq\cPgg$ (where $\cPq$ represents $\cPqu$ or $\cPqd$)
and $\cPqb\cPg \to \cPbst \to \cPqb\cPgg$ production by searching for resonances in $\GAMJET$ final states. The signal model includes
excited quarks with spin-$\frac{1}{2}$, and assumes a
compositeness scale that equals the mass of the resonance ($m_{\text{Res}}$). An assumption is also made that $f_{s}$, $f$, and $f'$ have identical
values~\cite{Baur1989kv, Bhattacharya:2009xg} and henceforth these will be referred to collectively as $f$. The data correspond to an integrated
luminosity of 35.9\fbinv collected by the CMS experiment in pp collisions at $\sqrt{s}=13\TeV$ at the CERN LHC, in 2016.

A final state with a photon and a jet is produced in the SM mainly through $\cPq\cPg\to\cPq\cPgg$, $\cPq\cPaq\to\cPg\cPgg$,
$\cPg\cPg\to\cPg\cPgg$, multijet, and
W/Z$+\cPgg$ processes. Among these, the main irreducible backgrounds are quark-gluon Compton scattering ($\cPq\cPg\to\cPq\cPgg$) and quark-antiquark
annihilation ($\cPq\cPaq\to\cPg\cPgg$). Although the probability for a jet to be reconstructed as a photon is ${\approx}10^{-4}$ to $10^{-3}$,
the cross section for multijet production is two to three orders of magnitude larger than that for the irreducible backgrounds,
depending on the $\pt$ of the jet~\cite{CMS:EGM-14-001}, making jet misidentification the second-largest source of background.
Electroweak production of $\PW$/$\PZ$+$\cPgg$,
where the W or Z boson decays to a pair of quark jets, contributes a very small fraction of the background due to its small production cross section.

This Letter provides a brief description of the CMS detector in Section 2. The main strategy used in selecting the events is
discussed in Section 3. Section 4 contains information about signal and background models, while Section 5 lists the
systematic uncertainties estimated in this analysis. The results of the study are presented in Section 6 and summarized in
Section 7.

\section{The CMS detector}
The central feature of the CMS detector is a superconducting solenoid of 6\unit{m} internal diameter, providing a magnetic field of 3.8\unit{T}.
Within the solenoid volume are a silicon pixel and strip tracker, a lead tungstate crystal electromagnetic calorimeter (ECAL), and a brass and
scintillator hadron calorimeter (HCAL), each composed of a barrel and two endcap sections. The very forward regions of the detector near the beam line is
covered by the forward calorimeters. Muons are measured in gas-ionization detectors embedded in the steel flux-return yoke outside the
solenoid. In the barrel section of the ECAL, an energy resolution of about 1\% is achieved for unconverted or late-converting photons in the tens
of \GeV energy range. The remaining barrel photons have a resolution of about 1.3\% up to pseudorapidity $\abs{\eta}$ = 1
rising to about 2.5\% at $\abs{\eta} = 1.4$~\cite{CMS:EGM-14-001}, where $\eta$ is defined as $-\ln[\tan({\theta}/2)]$, $\theta$ being the polar angle
of the cylindrical coordinates of the CMS detector.
In the endcaps, the resolution of unconverted or late-converting photons is about 2.5\%, while the remaining endcap photons
have a resolution between 3--4\%. When combining information from the entire detector, the jet energy resolution is
typically around 15\% at 10\GeV, 8\% at 100\GeV, and 4\% at 1\TeV. A more detailed description of the CMS detector, together with a definition of the coordinate system used and the relevant kinematic
variables, can be found in~\cite{Chatrchyan:2008zzk}.

The CMS experiment selects physics events using a two-tier trigger system, a hardware-based level-1 (L1) and a software-based high-level
trigger (HLT). The L1 trigger selects events of interest using information from the calorimeters and the muon system only, and reduces the readout
rate from the bunch crossing frequency of 40\unit{MHz} to below 100\unit{kHz}. The HLT system further decreases this rate to an
average of a few 100\unit{Hz} to a maximum of 1\unit{kHz}. The events selected
by the HLT are then reconstructed offline and used for analysis.
\section{Event selection}

Events are analyzed using a particle-flow (PF) algorithm~\cite{CMS-PRF-14-001}, which reconstructs and identifies each individual particle
with an optimized combination of information from the various elements of the CMS detector. The energy of photons is directly obtained from
the ECAL measurement, corrected for zero-suppression effects~\cite{CMS:EGM-14-001}. The energy of electrons is determined from a combination
of the electron momentum at the primary interaction vertex as determined by the tracker, the energy of the corresponding ECAL cluster, and
the energy sum of all bremsstrahlung photons spatially compatible with originating from the electron track. The energy of muons is obtained
from the curvature of the corresponding track. The energy of charged hadrons is determined from a combination of their momentum measured in the
tracker and the matching ECAL and HCAL energy deposits, corrected for zero-suppression effects and for the response function of the calorimeters
to hadronic showers. Finally, the energy of neutral hadrons is obtained from the corresponding corrected ECAL and HCAL energy.

The $\jets$ in each event are formed mainly from photons, charged, and neutral hadrons using the infrared- and collinear-safe anti-$\kt$
algorithm~\cite{Cacciari:2008gp}, with distance parameter $\Delta R = 0.4$ where $\Delta R = \sqrt{\smash[b]{{(\Delta \eta)}^{2} + {(\Delta \phi)}^{2}}}$, $\Delta \eta$ and $\Delta \phi$ being the pseudorapidity and azimuthal angle (in radians) difference between the jet axis and its constituents. Jet momenta and
energies are corrected to establish a uniform calorimetric response in $\eta$ and an absolute response in $\pt$ at the  particle level using
calibration constants~\cite{Khachatryan:2016kdb} obtained from simulation, test beam results, and pp collision data at $\sqrt{s}=13\TeV$.

The data sample used in this analysis consists of events that are selected by a photon trigger having a $\pt^{\cPgg}$ threshold
of 165\GeV and an additional condition on the ratio of the photon energy deposited in the HCAL to that in the ECAL (H/E), which is required to be less
than 10\%. The efficiency of the trigger used in the study has been evaluated separately using
samples collected with photon, muon, or jet triggers to account for possible biases in
trigger selection.
The trigger efficiencies measured in these samples are greater than 95\%
for $\pt^{\cPgg}> 200$\GeV, as measured offline.

In the offline selection, each event is required to have at least one
reconstructed primary vertex with at least four associated tracks, and lie within 24\unit{cm} along the $z$ direction
and within 2\unit{cm} in the transverse plane, from the nominal collision point.
The reconstructed vertex with the largest value of summed physics-object $\pt^2$
is taken as the primary pp interaction vertex.
The physics objects are the jets, clustered using the jet finding algorithm~\cite{Cacciari:2008gp,Cacciari2011ma} with the tracks assigned
to the vertex as inputs, and the associated missing transverse momentum, taken as the negative vector sum of the $\pt$ of those jets.

The photon identification~\cite{CMS:EGM-14-001} is based on requirements on H/E and shower profile of the photon.
The photon is isolated from identified electrons in the detector by requiring the absence of hits in the inner tracker layers near the photon direction.
The photon is also required to be well isolated from other photons and hadrons within a cone of $\Delta R = 0.3$ around
its axis. The photon must have $\PT^{\cPgg}> 200$\GeV\ and lie in the central barrel region
($|\eta^{\cPgg}|<1.4442$). Among the photons passing
the above criteria in each event, the one with the highest $\PT$ is selected to reconstruct the mass of the photon+jet system in the event. The isolation quantities are corrected for effects from
overlapping pp interactions (pileup) in the same or adjacent bunch crossings, by
subtracting the energy calculated from the mean energy density in the event, as computed using the $\FASTJET$ package~\cite{Cacciari2011ma}.
The photon identification and isolation criteria used in this analysis lead to a signal efficiency of $\sim$80\% with an
estimated background rejection of $\sim$90\%.

In order to be combined with a photon to form a resonance candidate, the selected $\jet$ must be separated from the chosen photon candidate by $\Delta R>0.5$
and satisfy the $\mathrm{tight}$ $\jet$ identification criteria~\cite{CMS-PAS-JME-16-003}. The $\jet$ identification criteria comprise requirements on
the number of constituents, and on the fraction of $\jet$ energy carried by each constituent type. The $\jet$ is required to be within
the region $|\eta^{\jet}|<2.4$ and must have a $\pt^{\jet}>170$\GeV. The angular separation between the
selected photon and jet is restricted by applying a requirement of $\Delta\eta\, (\gamma,\text{jet}) < 1.5$. This selection removes a large fraction of the
multijet background coming from non-isolated $\pi^{0}$s, without rejecting signal events, and thus enhances the signal-over-background ratio. If more than one $\jet$
candidate is present in the event, the $\jet$ with the highest \pt is used in the analysis. The selected events form the
``inclusive category'' for the search of light excited quarks.

\begin{figure}[htbp]
\centering
\includegraphics[width=0.49\textwidth]{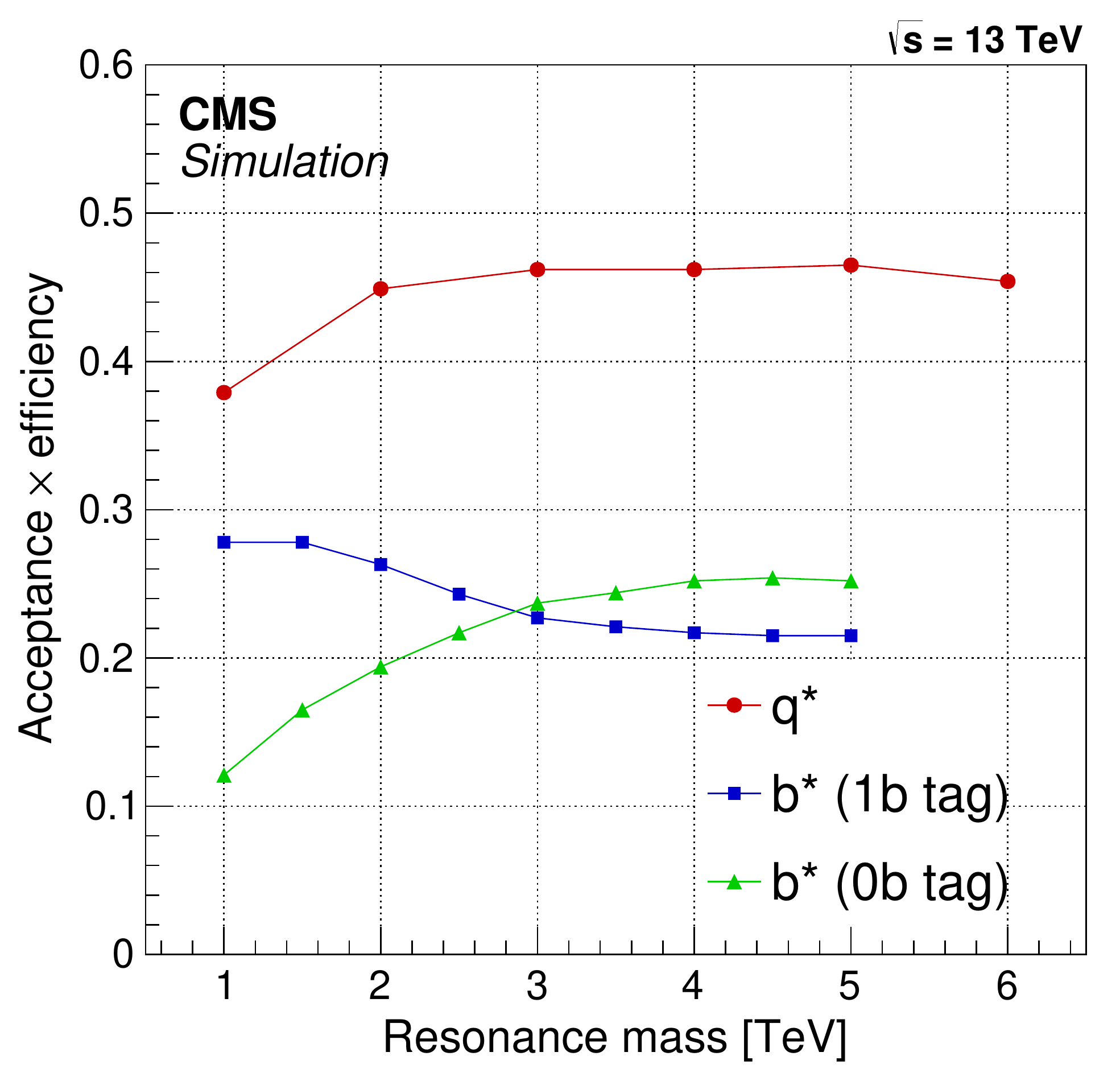}
 \caption{The product of acceptance and efficiency for $\cPqst$ and $\cPbst$ signals as a function of generated $m_{\cPqst}$ or $m_{\cPbst}$ mass, calculated using MC simulation.}
\label{fig:AccXEff}
\end{figure}

Jets originating from b quarks are identified using the combined secondary vertex v2 algorithm (CSVv2)~\cite{Sirunyan:2017ezt, Chatrchyan:2012jua}.
The algorithm combines the information from the primary vertex, impact parameters, and secondary vertices within the
jet using a neural network discriminator. The loose working point used in the analysis has $\sim$81\% b jet selection efficiency, $\sim$10\%
misidentification rate for light-quark and gluon jets, and $\sim$40\% misidentification rate for c quark jets~\cite{Sirunyan:2017ezt}. Depending on the outcome of the CSVv2 algorithm, a jet is tagged either as a b jet or a non b jet candidate. According to this tagging, for the $\cPbst$ analysis, the events are classified into ``1b tag''
and ``0b tag'' categories, corresponding to the selections with b jets and without b jets respectively.
Since the $\cPbst$ acceptance falls off slightly for 1b tag category at higher masses
(Fig.~\ref{fig:AccXEff}), the sensitivity of the search is improved by including the results from 0b tag category in the $\cPbst$ limit computation.

The above selection criteria are optimized for the best expected 95\% confidence level (CL) limits on the cross section versus mass of $\cPqst$ and $\cPbst$.

The efficiencies for assigning events to the 1b tag and 0b tag categories, determined from the Monte Carlo (MC) simulation, are corrected
using b tag scale factors (SFs), to take into account the observed differences between the b tagging efficiency of the CSVv2 tagger
applied to data and to MC simulation. The SFs are defined as ${\epsilon_{\text{data}}}/{\epsilon_{\text{MC}}}$,
where $\epsilon_{\text{data}}$ and $\epsilon_{\text{MC}}$ correspond to the b
tagging efficiencies of the CSVv2 algorithms in data and MC simulation, respectively.
These SFs have been measured using the techniques described in~\cite{Sirunyan:2017ezt}. 

The invariant mass of the selected $\GAMJET$ (\GAMbJET) system is required to be $m_{\cPgg+\jet} > 700\GeV$, to avoid the turn-on region due to
the requirements imposed on the kinematic properties of the trigger objects.
Fig.~\ref{fig:AccXEff} shows the total selection and reconstruction efficiency times acceptance for $\cPqst\to\cPq\cPgg$ and
$\cPbst\to\cPqb\cPgg$ processes. The acceptance times efficiency for the 1b tag category decreases with increasing mass owing to the decrease
in the efficiency of the track reconstruction and the resolution of the reconstructed track parameters with increasing $\pt$ of the $\jet$.

\section{Modeling signal and background}
The signal samples for $\cPqst$ and $\cPbst$ are simulated at leading order (LO)
with the \PYTHIA~8.212 event generator~\cite{Sjostrand:2014zea} for $f = 1.0$, 0.5 and 0.1 at different resonance masses
 in the range from 1 to 7\TeV at intervals of 1\TeV and from 1 to 5\TeV at intervals of 0.5\TeV, respectively.
The generated events are processed
through a full CMS detector simulation based on \GEANTfour~\cite{Agostinelli:2002hh}. The simulation uses the CUETP8M1 underlying event
tune~\cite{Khachatryan:2015pea,Skands:2014pea}, a renormalization and factorization scale corresponding to $\mu = \pt$ for the hard-scattered partons,
and NNPDF2.3LO parton distribution functions (PDFs)~\cite{Ball:2012cx}. The natural width of the resonance, at parton level, can be
approximated as $\Gamma\sim 0.03 f^{2} m_{\text{Res}}$~\cite{Baur1989kv}. The production cross section is also proportional to $f^2$. The
signals for intermediate mass points are interpolated at intervals of 50\GeV.

The \MGvATNLO v2.2.2 program~\cite{Alwall:2014hca} has been used to generate the $\GAMJET$ and $\PW$/$\PZ$+$\cPgg$ background MC samples at LO, with the showering
and hadronization carried out
by the \PYTHIA~8.212 program. A double counting of the partons
generated with \MADGRAPH and those with \PYTHIA is removed using the MLM~\cite{Alwall:2007fs} matching scheme. The multijet MC events are generated using \PYTHIA~8.212 event generator. The same event reconstruction is employed in data and MC simulations.
However, the background is evaluated from data, and the MC simulation is used only for the optimization of the event selection.
The invariant mass distribution of the SM $\GAMJET$ background falls smoothly and can be described by an analytic function.

\begin{figure}[htb]
\centering
\includegraphics[width=0.49\textwidth]{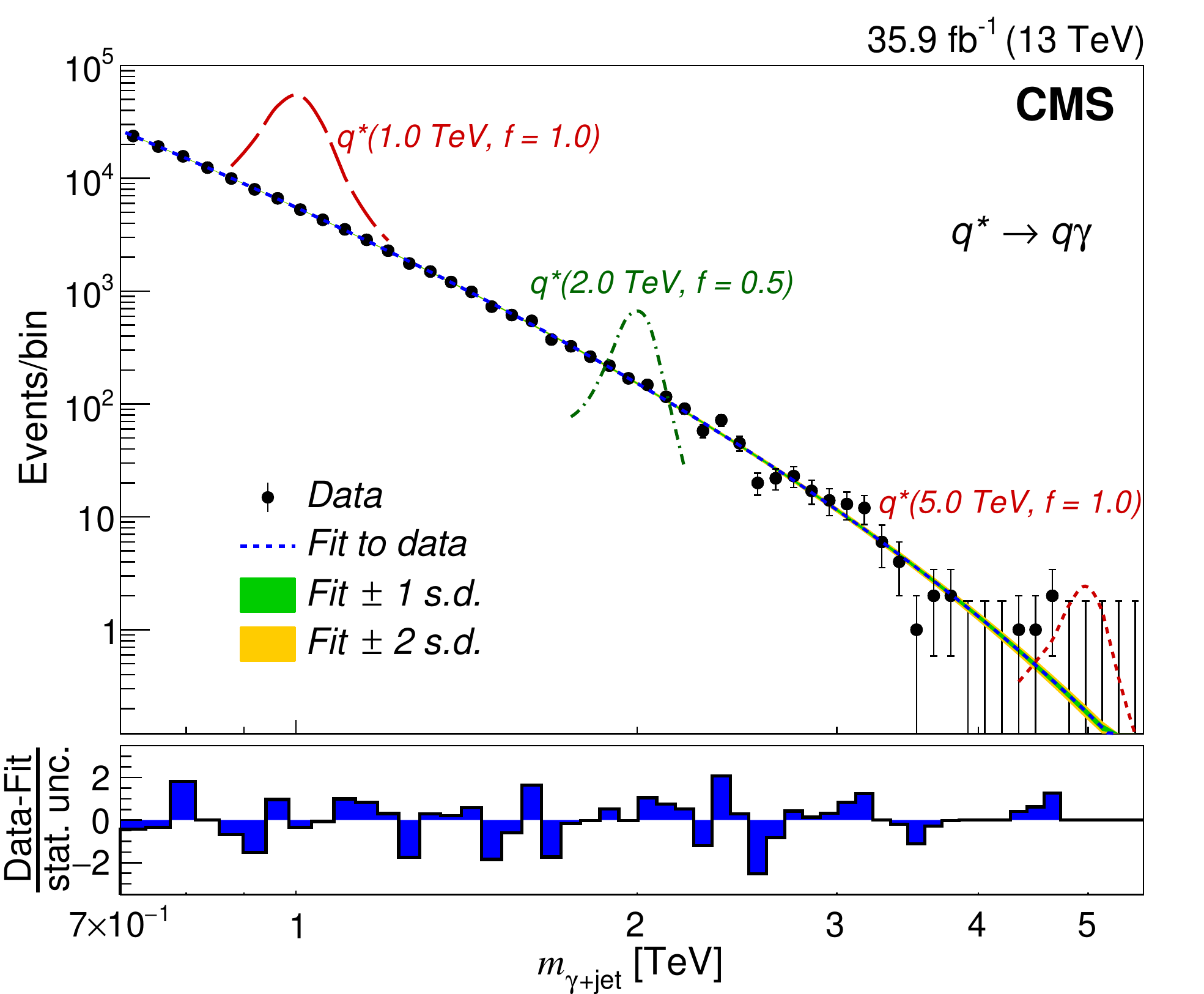}
\caption{The $\GAMJET$ invariant mass distribution in data (black points) for the inclusive category used for the $\cPqst$ analysis,
after final selection. The result of the fit to the data using the parametrization defined in Eq.~(\ref{eq:bkgfit}) is
shown by the blue dashed curve with associated bands indicating the uncertainty. The bin-by-bin pull, (Data-Fit)/(stat. unc.), where the denominator refers to the statistical uncertainty in data, is also presented. The green and yellow bands correspond to 1 and 2 standard deviations, respectively from the mean value.
Simulations of excited quark signals representing the expected excess of signal events over the background are shown for the mass values of 1.0 and 5.0\TeV for
$f = 1.0$, and 2.0\TeV for $f = 0.5$.}
\label{fig:gjMass}
\end{figure}

\begin{figure}[htbp]
\centering
\includegraphics[width=0.49\textwidth]{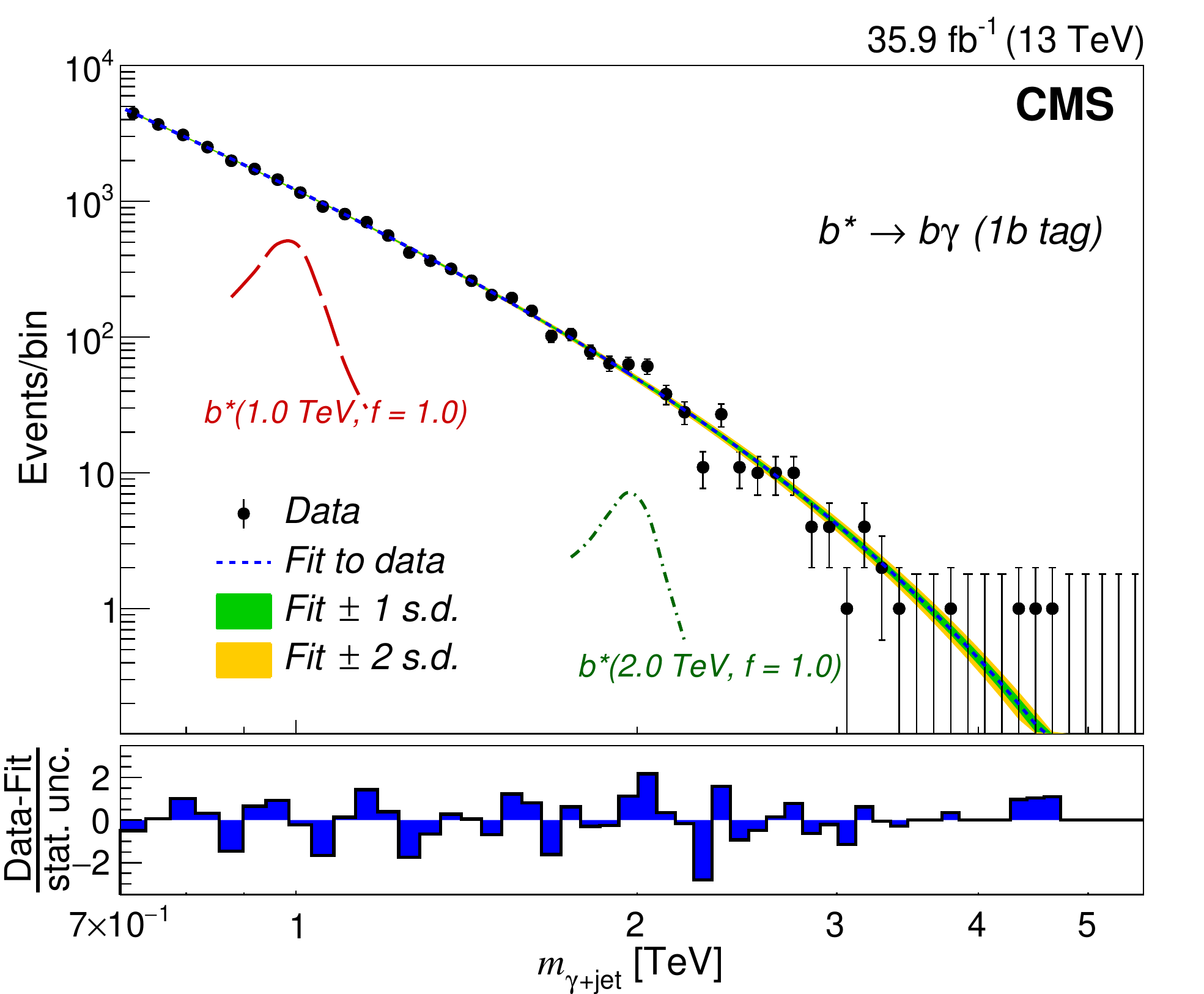}
\includegraphics[width=0.49\textwidth]{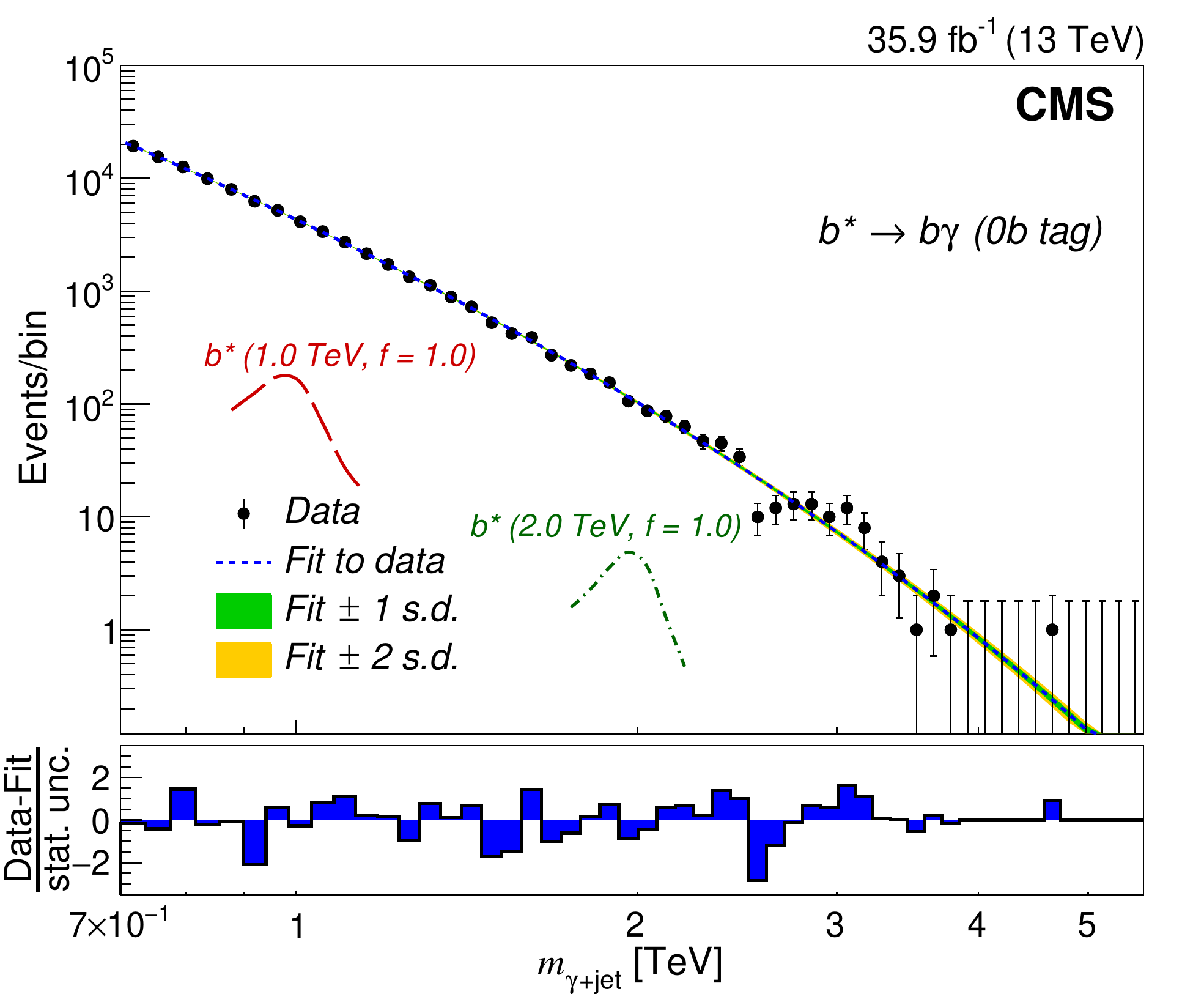}
\caption{The $\GAMbJET$ invariant mass distribution in data (black points) used for the $\cPbst$ analysis, after final selection for (\cmsLeft) 1b tag category and (\cmsRight) 0b tag category. The result of the fit to the data using the parametrization defined in Eq.~(\ref{eq:bkgfit}) is
shown by the blue dashed curve with associated bands indicating the uncertainty. The bin-by-bin pull, (Data-Fit)/(stat. unc.), where the denominator refers to the statistical uncertainty in data, is also presented. The green and yellow bands correspond to 1 and 2 standard deviations, respectively from the mean value.
Simulations of excited b quark signals representing the expected excess of signal events over the background are shown for the 1b and 0b tag categories
for the mass values of 1.0 and 2.0\TeV for $f = 1.0$.}
\label{fig:gjMass_1tag}
\end{figure}

The inclusive invariant mass distribution and the distributions for 1b tag and 0b tag categories, expressed in \TeV,  are shown in Figs.~\ref{fig:gjMass}
and~\ref{fig:gjMass_1tag}, respectively. The binning is chosen to have a bin width approximately equal to the expected $\GAMJET$ mass
resolution, which varies from about 4.5\% at a mass of 1\TeV to 3.3\% at 6\TeV. These distributions are modeled using an
empirical parametrization that has been used widely in similar previous searches~\cite{Aaltonen:2008dn, Khachatryan:2014aka, Khachatryan:2015dcf, Aad:2015ywd}:
\begin{equation}
\frac{\rd\sigma}{\rd{}m} = \frac{P_{0}(1-m/\sqrt{s})^{P_{1}}}{(m/\sqrt{s})^{P_{2}+P_{3}\ln(m/\sqrt{s})}}
\label{eq:bkgfit}
\end{equation}
where $\sqrt{s} = 13\TeV$ and $P_{0}$, $P_{1}$, $P_{2}$, and $P_{3}$ are four parameters used to describe the background
distribution and its normalization. The order of the function has been chosen by performing Fisher tests~\cite{10.2307/2340521}, with a cut-off
p-value of 0.05. The function is found to be in good agreement with data with a $\chi^{2}/\textrm{ndf}$ = $40.7/41.4$.
The highest invariant mass event observed in data has $m_{\cPgg+\jet}$ of 4.6\TeV with a b-tagged $\jet$, and thus
belongs to both the inclusive and 1b tag categories.

In order to examine the presence of a possible systematic bias due to the choice of background fitting function, tests are performed
using alternate functional forms. These alternative expressions are polynomial functions that also provide adequate descriptions of the data.
To perform these tests, an invariant mass distribution of the SM background is
obtained from MC simulation.
This invariant mass distribution is fitted with alternate test functions and the results of the fit, considered as the truth model, are used to generate
a large number of pseudo-data samples that have bin-to-bin statistical fluctuations similar to those of the data.
A signal with a cross section close to
the expected sensitivity is also injected in the pseudo-data distributions. These distributions are then fitted using the default
background function along with a signal model, and the signal cross section is extracted. Pull distributions defined
as the difference between the true and extracted signal cross sections divided by the estimated statistical uncertainty, for the obtained
signal cross sections are constructed. The deviation from zero, of the mean in the pull
distribution, is a measure of the bias present in the model. The pull distributions for $\cPqst$ and $\cPbst$ modeling over the studied mass range
are found to be consistent with normalized Gaussian forms with medians deviating
by no more than 0.5 from zero,
and widths consistent with unity for the full mass range. When added in quadrature with the statistical uncertainty, the bias uncertainty
is found to contribute approximately 10\% of the total.
Therefore, it is concluded that the systematic uncertainty associated with the
choice of the parametric function is negligible, and the statistical uncertainty of
the fit is the only uncertainty in the background prediction that needs to be
considered.

\section{Systematic uncertainties}
The dominant sources of the systematic uncertainties affecting the $\cPqst$ and $\cPbst$ signals are summarized in Table~\ref{Table:SysUnc}.
\begin{table*}[h!t]
\centering
\topcaption{Summary of the dominant sources of uncertainties and their effect on
  the signal yield.}
\begin{tabular}{lc}
Source   & Effect on the signal yield(\%) \\
\hline
Integrated luminosity                            & 2.5      \\
Jet energy scale                      & $\sim$1  \\
Jet energy resolution                 & 0.2--0.4  \\
Photon energy scale                   & $\sim$0.6  \\
Photon energy resolution              & 0.2--0.4   \\
Pileup                                 & 1--2  \\
Photon ID efficiency                & $\sim$2  \\
Trigger efficiency              & $\sim$5       \\
Signal interpolation            & 0.5--1  \\
PDF choice                       &  1.5--3   \\
b tag SF (only \cPbst)         & $\sim$1       \\
b tag SF normalization (only \cPbst)  & $\sim$2   \\
\end{tabular}
   \label{Table:SysUnc}
\end{table*}

The uncertainties in the jet energy scale and jet energy resolution~\cite{Khachatryan:2016kdb} affect both the
signal yield and its distribution. The size of the effect is determined by varying the four-momenta of the jets by the corresponding uncertainties
and repeating the full analysis with the modified quantities.

The systematic uncertainties in the photon energy scale and resolution, and photon identification
efficiency are derived from $\cPZ\to\EE$ events. The  uncertainty
in the photon energy scale is found to be about 1\% and it includes the uncertainty in the extrapolation to higher \pt, beyond the reach of
the $\cPZ\to\EE$ control samples~\cite{CMS:EGM-14-001}. The uncertainty in the photon identification is estimated to be around 2\%.
Also, a systematic uncertainty of 5\% has been
included to account for the precision of the photon trigger efficiency
measurement. The effect of the b tagging scale-factor uncertainty on the distribution of the signal
is evaluated to be around 1\% while on the normalization, the effect is around 2\%.
The method used to interpolate the signal distributions from the generated distributions is assigned an uncertainty of 0.5--1.0\%,
which accounts for the difference between the generated and interpolated signals.
The PDF uncertainty affects the signal acceptance by 1.5--3.0\% for both $\cPqst$ and $\cPbst$ quarks and is evaluated using PDF4LHC
recommendations~\cite{0954-3899-43-2-023001}.

The uncertainties in the measurement of the integrated luminosity (2.5\%)~\cite{CMS:2017LUM} and pileup description (1\%) affect the overall signal yield.
The uncertainty in the background estimate is accounted for in the fit by
varying the parameter values within their respective uncertainties, with no additional constraints.

\section{Results}
In the mass region studied, no significant excess has been observed. We use the $\GAMJET$ invariant mass spectra (Figs.~\ref{fig:gjMass},~\ref{fig:gjMass_1tag}),
the background parametrization, and the $\cPqst$ and $\cPbst$ theoretical predictions
to set 95\% CL upper limits on the production cross section of $\cPqst$ and $\cPbst$ decaying to $\cPq\cPgg$ and $\cPqb\cPgg$, respectively.

The modified frequentist CL$_\mathrm{s}$ method~\cite{Junk:1999kv,Read:2002hq} in the asymptotic approximation~\cite{Cowan:2010js} is utilized to set upper limits
on signal cross sections. The asymptotic approximation is found to be in good agreement with the full CL$_\mathrm{s}$ approximation over the entire mass range.
In order to evaluate limits, a likelihood function is constructed that is the product of the Poisson likelihoods of all the bins in the
distribution.
The systematic uncertainties in the signal are implemented in terms of nuisance parameters with Gaussian and log-normal constraints.
The uncertainty due to the background parametrization is found to have the largest impact and is quantified by considering the effect of changing the
parameters from their central values by their estimated $\pm$1 sigma uncertainties.
We calculate limits by evaluating the likelihood independently at successive values of resonance mass from 1 to 6\TeV
for $\cPqst$, and 1 to 5\TeV for $\cPbst$ in steps of 50\GeV. The cross section limits are not evaluated below 1\TeV, because of
uncertainties in the signal efficiency associated with the invariant mass selection, $m_{\cPgg+\jet} > 700\GeV$.

\begin{figure}[htb]
\centering
\includegraphics[width=0.49\textwidth]{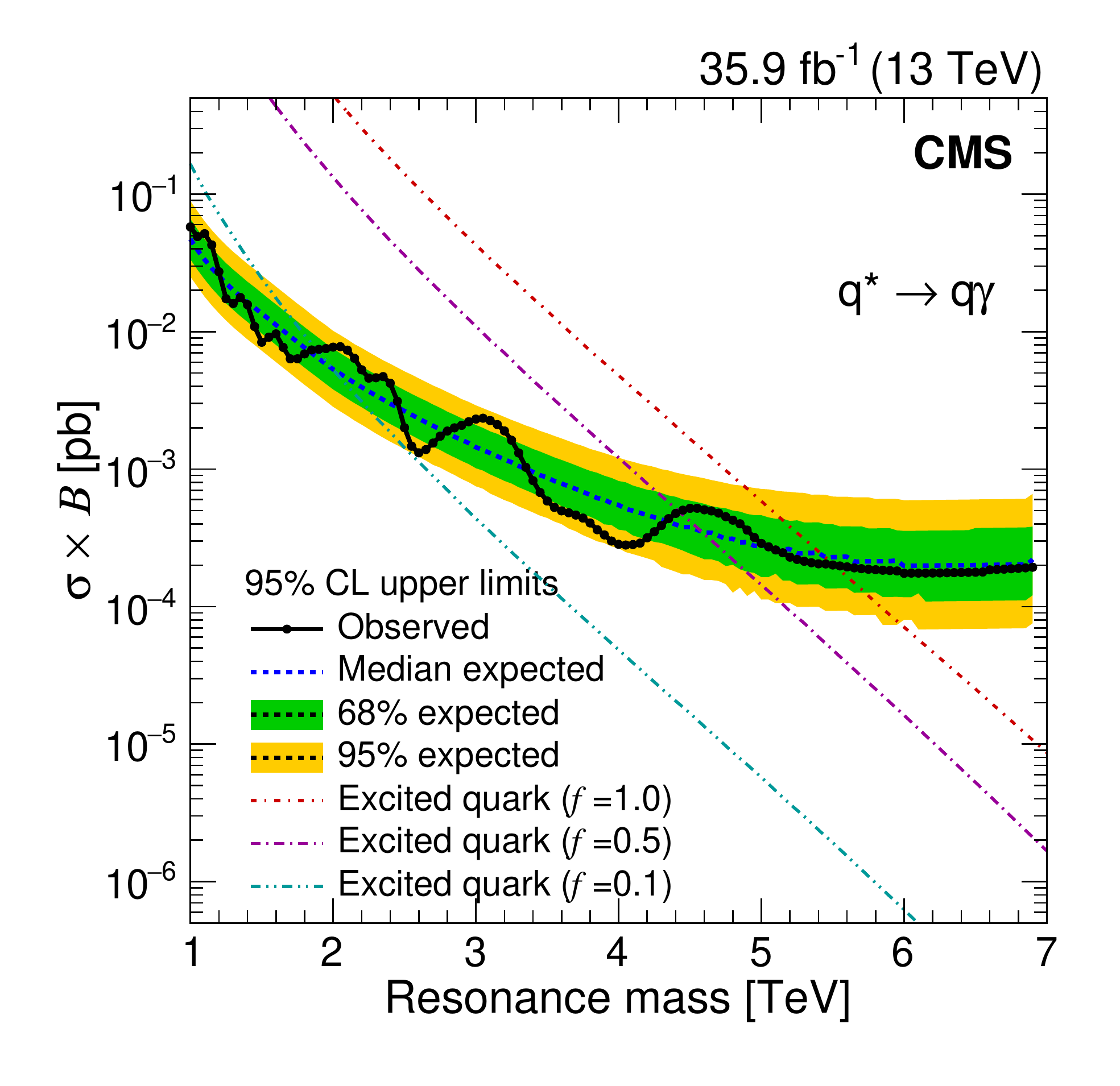}
 \caption{The observed and expected upper limits at 95\% CL on $\sigma\,\mathcal{B}$ as a function of the mass of the excited quark, for $f = 1.0$.
 The limits are compared with theoretical predictions for excited quark production for three couplings. The inner (green) band and the
outer (yellow) band indicate the regions containing 68 and 95\%, respectively, of
the mean limits under the background-only hypothesis.}
\label{fig:qstarlimit}
\end{figure}

\begin{figure}[htb]
\centering
\includegraphics[width=0.49\textwidth]{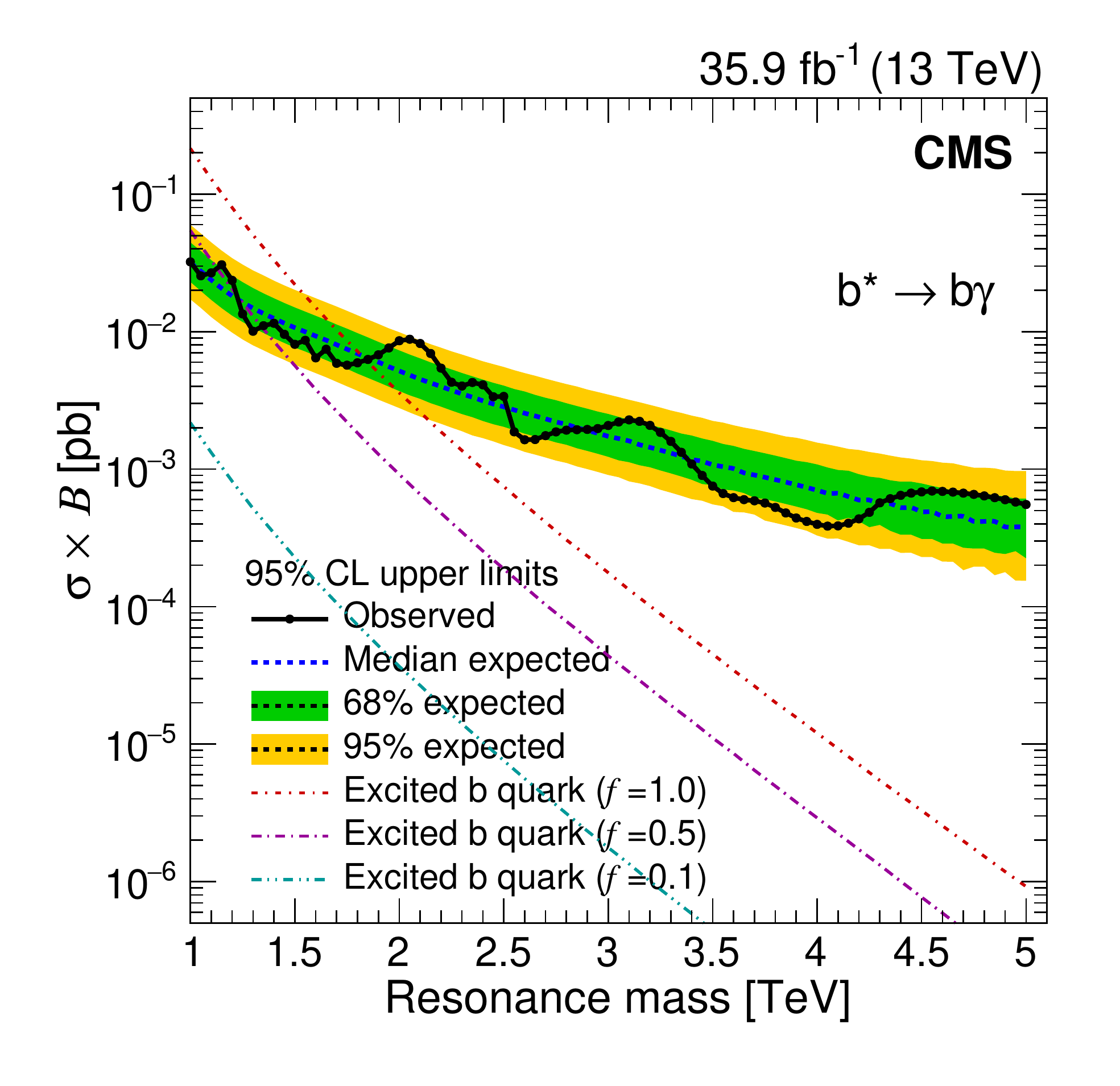}
 \caption{The observed and expected upper limits at 95\% CL on $\sigma\,\mathcal{B}$ as a function of the mass of the excited b quark, for $f = 1.0$.
 The limits are compared with theoretical predictions for excited b quark production for three couplings. The inner (green) band
and the outer (yellow) band indicate the regions containing 68 and 95\%,
respectively, of the mean limits under the background-only hypothesis.}
\label{fig:bstarlimits_1tag}
\end{figure}

\begin{figure}[htb]
\centering
\includegraphics[width=0.49\textwidth]{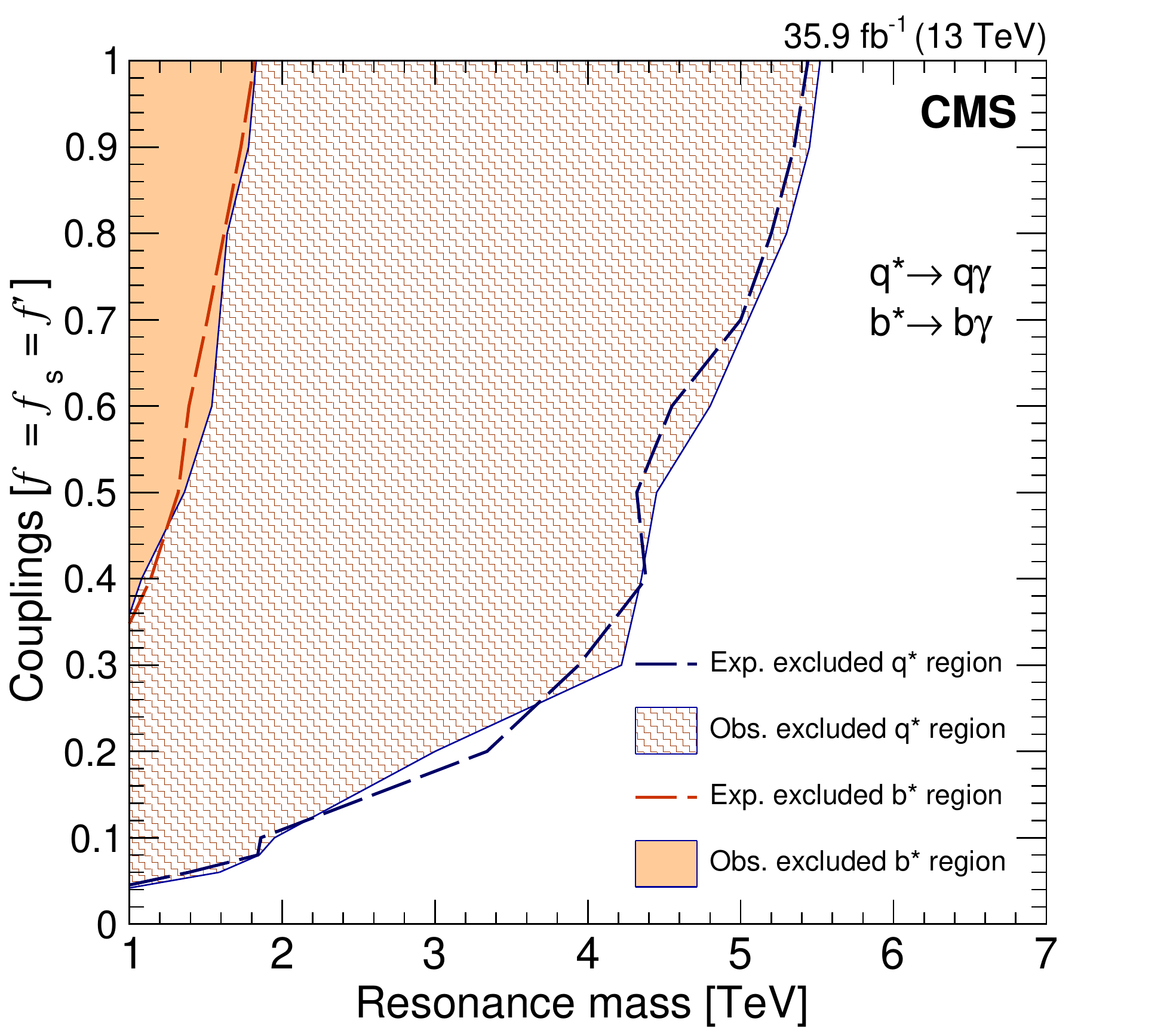}
\caption{The observed and expected regions excluded at 95\% CL for $\cPqst$ and $\cPbst$ production and decay, as a function of $m_{\cPqst}$, $m_{\cPbst}$ and $f$.}
\label{fig:2dcoup_all}
\end{figure}

In order to evaluate limits for $\cPbst$, likelihoods for 1b and 0b tag categories are combined together.
The observed and expected mass limits for $\cPqst$ and $\cPbst$ are computed at 95$\%$ CL.\ The results are presented in terms of limits
 on the product of the cross section ($\sigma$) and branching fraction ($\mathcal{B}$).
The cross section upper limits are compared to the LO theoretical predictions, for all the three couplings, to estimate the lower mass limit on
excited quarks. In Figs.~\ref{fig:qstarlimit} and~\ref{fig:bstarlimits_1tag}, the experimental limits for $f=1.0$ are shown for $\cPqst$ and $\cPbst$,
respectively, with the theoretical predictions for the different couplings overlaid. There is a small dependence of $\sigma \times \mathcal{B}$ on $f$,
of the order of $10\%-20\%$, which is taken into account correctly when extracting the mass limits.
Observed lower bounds of 5.5 and 1.8\TeV are obtained for $\cPqst$ and $\cPbst$, respectively, for $f=1.0$. The
corresponding expected mass limits obtained are 5.4 (1.8)\TeV for $\cPqst$ ($\cPbst$).
The variation of the excluded mass as a function of the coupling strength, obtained by interpolating the efficiencies for
three signal MC samples corresponding to $f = 1.0$, 0.5, 0.1, is shown for $\cPqst$ and $\cPbst$in Fig.~\ref{fig:2dcoup_all}.
This result can also be interpreted in terms of the
ratio of the resonance mass and $\Lambda$, \ie, if we relax the assumption of $\Lambda$ = $m_{\text{Res}}$, the excited quark production cross section
is proportional to $f$ as well as $m_{\text{Res}}$/$\Lambda$.

\section{Summary}
A search has been presented for excited states of light and b quarks in $\GAMJET$ final states, using data corresponding to an integrated luminosity of
35.9\fbinv, collected at $\sqrt{s} = 13\TeV$. Upper limits at the 95\% confidence level are placed on the product of production cross section
and decay branching fraction for the presence of $\cPqst$ and $\cPbst$ excited quarks in $\GAMJET$ final states.
Comparing these upper limits with theoretical predictions, excited light quarks within the mass range $1.0 < m_\cPqst < 5.5\TeV$
and excited b quarks within the mass range $1.0 < m_\cPbst < 1.8\TeV$ are excluded at 95\% confidence level, assuming standard model like coupling strengths.
These are the most sensitive limits for $\cPqst$ and $\cPbst$ searches in the $\GAMJET$ final
states. In addition, the search for excited b quarks is the first to be presented in
any final state at $\sqrt{s} = 13\TeV$.

\begin{acknowledgments}
We congratulate our colleagues in the CERN accelerator departments for the excellent performance of the LHC and thank the technical and administrative staffs at CERN and at other CMS institutes for their contributions to the success of the CMS effort. In addition, we gratefully acknowledge the computing centers and personnel of the Worldwide LHC Computing Grid for delivering so effectively the computing infrastructure essential to our analyses. Finally, we acknowledge the enduring support for the construction and operation of the LHC and the CMS detector provided by the following funding agencies: BMWFW and FWF (Austria); FNRS and FWO (Belgium); CNPq, CAPES, FAPERJ, and FAPESP (Brazil); MES (Bulgaria); CERN; CAS, MOST, and NSFC (China); COLCIENCIAS (Colombia); MSES and CSF (Croatia); RPF (Cyprus); SENESCYT (Ecuador); MoER, ERC IUT, and ERDF (Estonia); Academy of Finland, MEC, and HIP (Finland); CEA and CNRS/IN2P3 (France); BMBF, DFG, and HGF (Germany); GSRT (Greece); OTKA and NIH (Hungary); DAE and DST (India); IPM (Iran); SFI (Ireland); INFN (Italy); MSIP and NRF (Republic of Korea); LAS (Lithuania); MOE and UM (Malaysia); BUAP, CINVESTAV, CONACYT, LNS, SEP, and UASLP-FAI (Mexico); MBIE (New Zealand); PAEC (Pakistan); MSHE and NSC (Poland); FCT (Portugal); JINR (Dubna); MON, ROSATOM, RAS, RFBR and RAEP (Russia); MESTD (Serbia); SEIDI, CPAN, PCTI and FEDER (Spain); Swiss Funding Agencies (Switzerland); MST (Taipei); ThEPCenter, IPST, STAR, and NSTDA (Thailand); TUBITAK and TAEK (Turkey); NASU and SFFR (Ukraine); STFC (United Kingdom); DOE and NSF (USA).

\hyphenation{Rachada-pisek} Individuals have received support from the Marie-Curie program and the European Research Council and Horizon 2020 Grant, contract No. 675440 (European Union); the Leventis Foundation; the A. P. Sloan Foundation; the Alexander von Humboldt Foundation; the Belgian Federal Science Policy Office; the Fonds pour la Formation \`a la Recherche dans l'Industrie et dans l'Agriculture (FRIA-Belgium); the Agentschap voor Innovatie door Wetenschap en Technologie (IWT-Belgium); the Ministry of Education, Youth and Sports (MEYS) of the Czech Republic; the Council of Science and Industrial Research, India; the HOMING PLUS program of the Foundation for Polish Science, cofinanced from European Union, Regional Development Fund, the Mobility Plus program of the Ministry of Science and Higher Education, the National Science Center (Poland), contracts Harmonia 2014/14/M/ST2/00428, Opus 2014/13/B/ST2/02543, 2014/15/B/ST2/03998, and 2015/19/B/ST2/02861, Sonata-bis 2012/07/E/ST2/01406; the National Priorities Research Program by Qatar National Research Fund; the Programa Severo Ochoa del Principado de Asturias; the Thalis and Aristeia programs cofinanced by EU-ESF and the Greek NSRF; the Rachadapisek Sompot Fund for Postdoctoral Fellowship, Chulalongkorn University and the Chulalongkorn Academic into Its 2nd Century Project Advancement Project (Thailand); the Welch Foundation, contract C-1845; and the Weston Havens Foundation (USA). \end{acknowledgments}

\bibliography{auto_generated}

\cleardoublepage \appendix\section{The CMS Collaboration \label{app:collab}}\begin{sloppypar}\hyphenpenalty=5000\widowpenalty=500\clubpenalty=5000\textbf{Yerevan Physics Institute,  Yerevan,  Armenia}\\*[0pt]
A.M.~Sirunyan, A.~Tumasyan
\vskip\cmsinstskip
\textbf{Institut f\"{u}r Hochenergiephysik,  Wien,  Austria}\\*[0pt]
W.~Adam, F.~Ambrogi, E.~Asilar, T.~Bergauer, J.~Brandstetter, E.~Brondolin, M.~Dragicevic, J.~Er\"{o}, M.~Flechl, M.~Friedl, R.~Fr\"{u}hwirth\cmsAuthorMark{1}, V.M.~Ghete, J.~Grossmann, J.~Hrubec, M.~Jeitler\cmsAuthorMark{1}, A.~K\"{o}nig, N.~Krammer, I.~Kr\"{a}tschmer, D.~Liko, T.~Madlener, I.~Mikulec, E.~Pree, N.~Rad, H.~Rohringer, J.~Schieck\cmsAuthorMark{1}, R.~Sch\"{o}fbeck, M.~Spanring, D.~Spitzbart, W.~Waltenberger, J.~Wittmann, C.-E.~Wulz\cmsAuthorMark{1}, M.~Zarucki
% ...
\vskip\cmsinstskip
% ...
1:~~Also at Vienna University of Technology, Vienna, Austria\\
% ...

\end{sloppypar}
\end{document}